\documentclass[a4paper]{PoS}
\usepackage{bm,amsmath,amssymb}
\usepackage{graphicx}
\usepackage{subfigure}
\usepackage{color}
\usepackage{soul}
\setulcolor{blue}
\setstcolor{red}
\sethlcolor{yellow}

\newcommand \beq {\begin{equation}}
\newcommand \eeq {\end{equation}}
\newcommand \beqa {\begin{eqnarray}}
\newcommand \eeqa {\end{eqnarray}}
\newcommand \lrnd {\left(}
\newcommand \rrnd {\right)}

\newcommand \lang {\left\langle}
\newcommand \rang {\right\rangle}

\newcommand \MeV {\,\text{MeV}}
\newcommand \GeV {\,\text{GeV}}

\newcommand \hmu {\hat{\mu}}
\title{Charge Fluctuations as Thermometer for Heavy-Ion Collisions}

\ShortTitle{Charge Fluctuations as Thermometer for Heavy-Ion Collisions}

\author{\speaker{Mathias Wagner} (for BNL-Bielefeld Collaboration)\\
       Fakult\"at f\"ur Physik, Universit\"at Bielefeld, D-33615 Bielefeld, Germany\\
       E-mail: \email{mwagner@physik.uni-bielefeld.de}}

\abstract{We present a determination of freeze-out conditions in heavy-ion collisions based on ratios of cumulants of net electric charge fluctuations obtained from lattice QCD. These ratios can reliably be calculated for a wide range of chemical potential values by using a next-to-leading order Taylor series expansion around the limit of vanishing baryon, electric charge and strangeness chemical potentials. We first determine the strangeness and electric charge chemical potentials that characterize the conditions in heavy ion collisions at RHIC and LHC. We then show that a comparison of lattice QCD results for ratios of up to third order cumulants of electric charge fluctuations with experimental results allows us to extract the freeze-out baryon chemical potential and the freeze-out temperature. We apply our method to preliminary data of the STAR and PHENIX collaborations.}

\FullConference{31st International Symposium on Lattice Field Theory LATTICE 2013\\
		 July 29 - August 3, 2013\\
		 Mainz, Germany}

\begin{document}

\section{Introduction}

The exploration of the phase structure of strongly-interacting matter is one of the major goals of the experimental programs at the Relativistic Heavy-Ion Collider (RHIC) as well as for future experiments at the upcoming FAIR and NICA facilities. 
To interpret these explorations the measurement of observables that can  be connected to first-principles theoretical investigations is favorable. The  fluctuations of conserved charges, e.g., baryon number (B), electric charge (Q) and strangeness (S) have turned out to meet this criteria.  
The experimentally measured fluctuations stem from the time when hadrons reappeared and reflect the conditions at the chemical freeze-out. To reflect any signals of critical behavior the freeze-out must occur close to the QCD phase boundary. This phase boundary is quite reliably known for small chemical potential from first-principle Lattice QCD simulations~\cite{strange,Tc,curv}. The freeze-out temperature and chemical potential, however, are traditionally obtained from fits of the measured hadrons yields to statistical hadronization models~\cite{pbm}. Although
 this successful description  suggests that these freeze-out conditions can indeed be characterized by a freeze-out temperature ($T^f$) and baryon chemical potential ($\mu_B^f$) one would clearly prefer a determination of these important parameters based on first-principles. 

Here we will discuss such an approach which relies on the calculation of the fluctuations of the conserved charges from LQCD. These fluctuations are commonly discussed in terms of the generalized susceptibilities
\beq
\chi_{mn}^{XY} = \left. \frac{\partial^{(m+n)} [p(\hmu_X,\hmu_Y)/T^4]} 
{\partial \hmu_X^m \partial \hmu_Y^n} \right|_{\vec{\mu}=0}
\ ,
\label{eq:susc}
\eeq
where $\vec{\mu}=(\mu_B,\mu_S,\mu_Q)$ are respectively the baryon number, strangeness
and electric charge chemical potentials and $X,Y=B,S,Q$. 
We use the
notations $\chi_{0n}^{XY}\equiv\chi_n^Y$ and $\chi_{m0}^{XY}\equiv\chi_m^X$.  These
generalized susceptibilities are related to the cumulants, such as the mean ($M_X$),
variance ($\sigma_X$), skewness ($S_X$) and kurtosis ($\kappa_X$), of the
fluctuations of the conserved charge. For example--- $VT^3\chi_1^Q=\lang
N_Q\rang=M_Q$, $VT^3\chi_2^Q=\lang(\delta N_Q)^2\rang=\sigma_Q^2$,
$VT^3\chi_3^Q=\lang(\delta N_Q)^3\rang=\sigma_Q^3S_Q$ and $VT^3\chi_4^Q=\lang(\delta
N_Q)^4\rang - 3\lang(\delta N_Q)^2\rang^2=\sigma_Q^4\kappa_Q$; $V$ being the
volume, $T$ the temperature and $N_X$ the net charge with $\delta N_X = N_X
-\lang N_X\rang$.  
The generalized susceptibilities can be calculated using standard lattice techniques for a wide range of chemical potentials by using a next-to-leading order Taylor series expansion. Here we rely on data calculated on lattices with temporal extent $N_\tau=6,8,12$ using staggered fermions (highly-improved staggered quarks) with $2+1$ flavors.
Details of the LQCD calculations presented here can be found in~\cite{strange,Tc,freeze,hrg}. 

\section{Strangeness and electric charge chemical potentials}
\begin{figure}
\subfigure[]{\label{fig:qi} 
\includegraphics[width=0.313\textwidth]{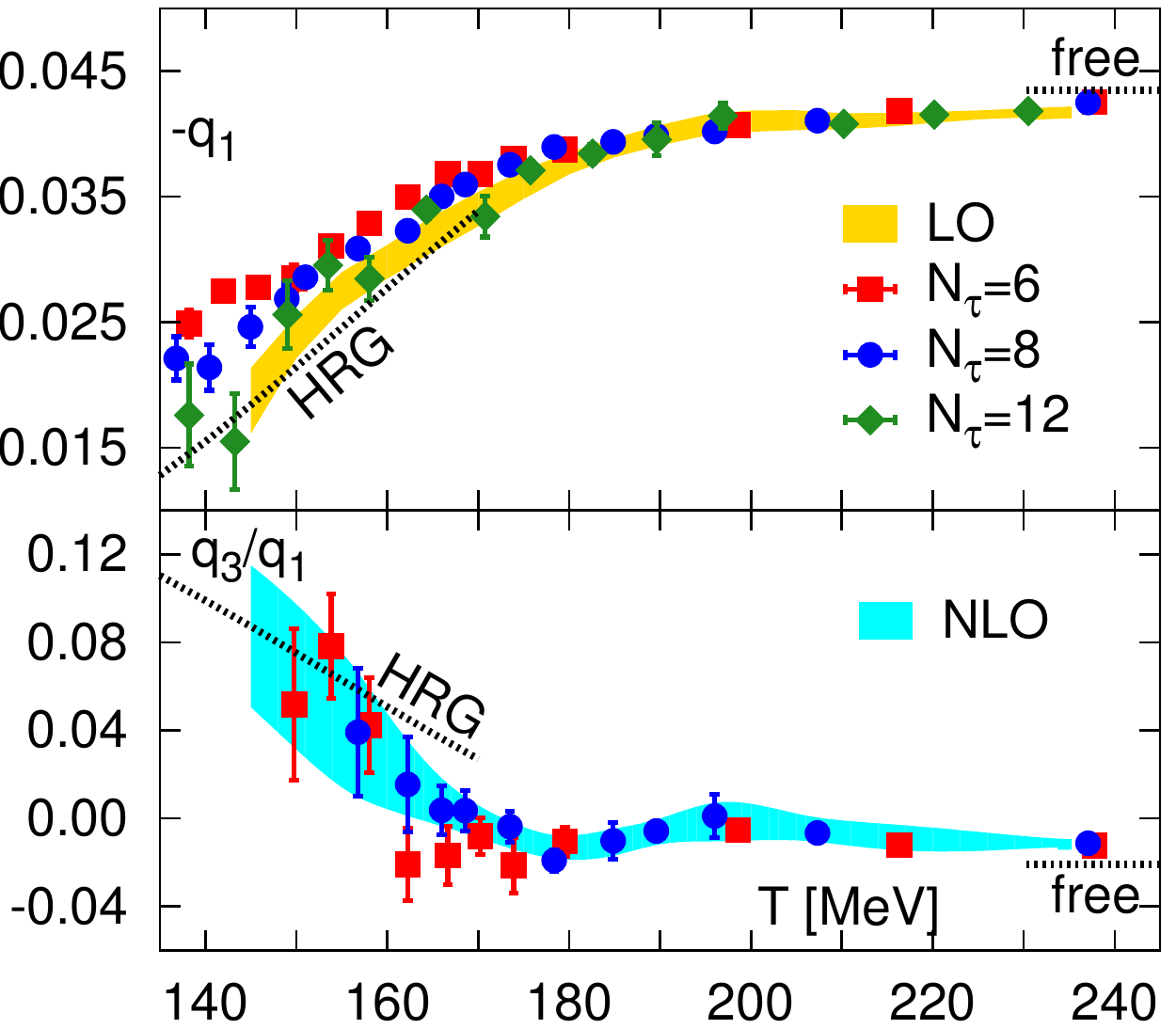} }
\subfigure[]{ \label{fig:si} 
\includegraphics[width=0.313\textwidth]{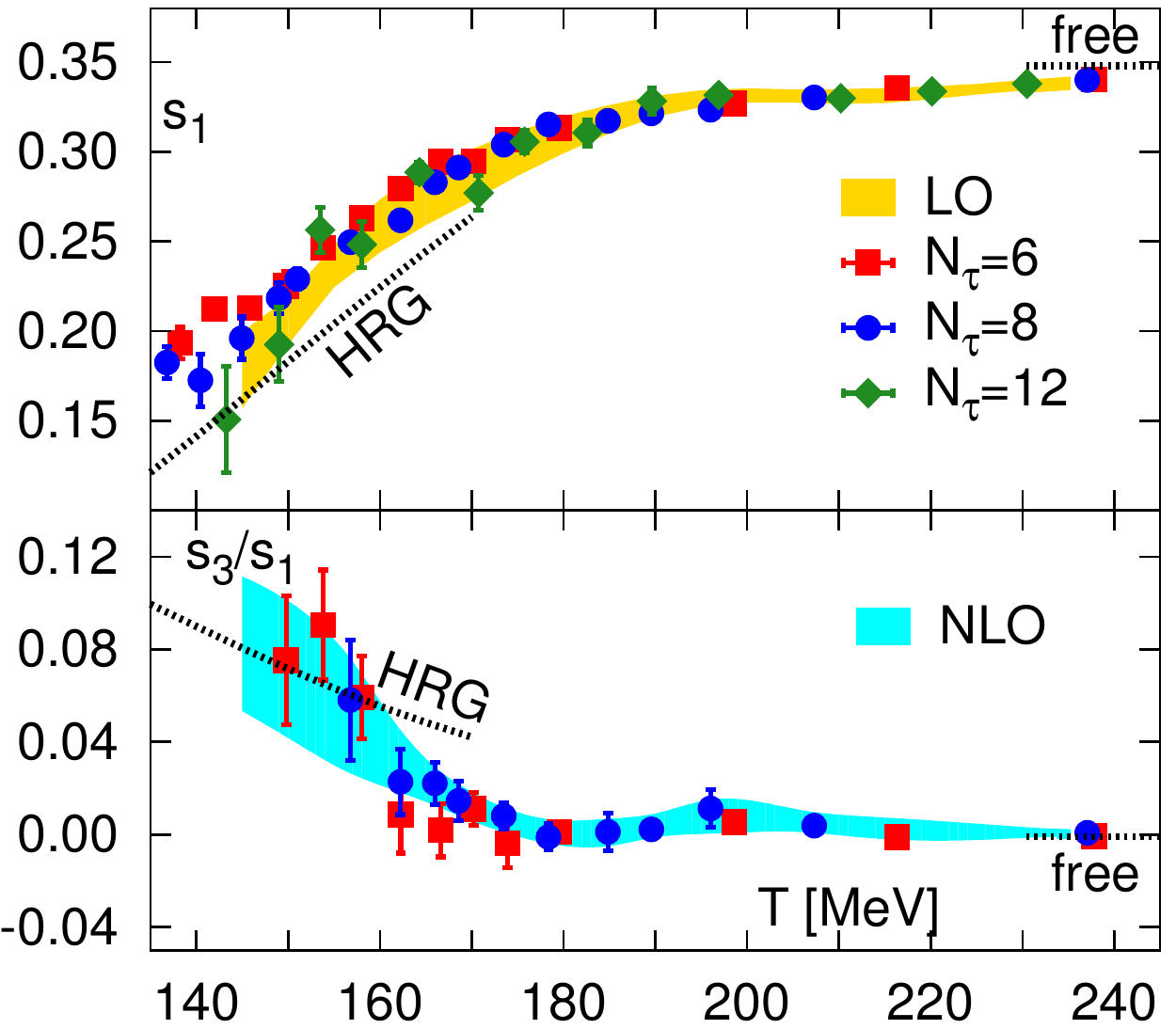} }
\subfigure[]{ \label{fig:muQS} 
\includegraphics[width=0.313\textwidth]{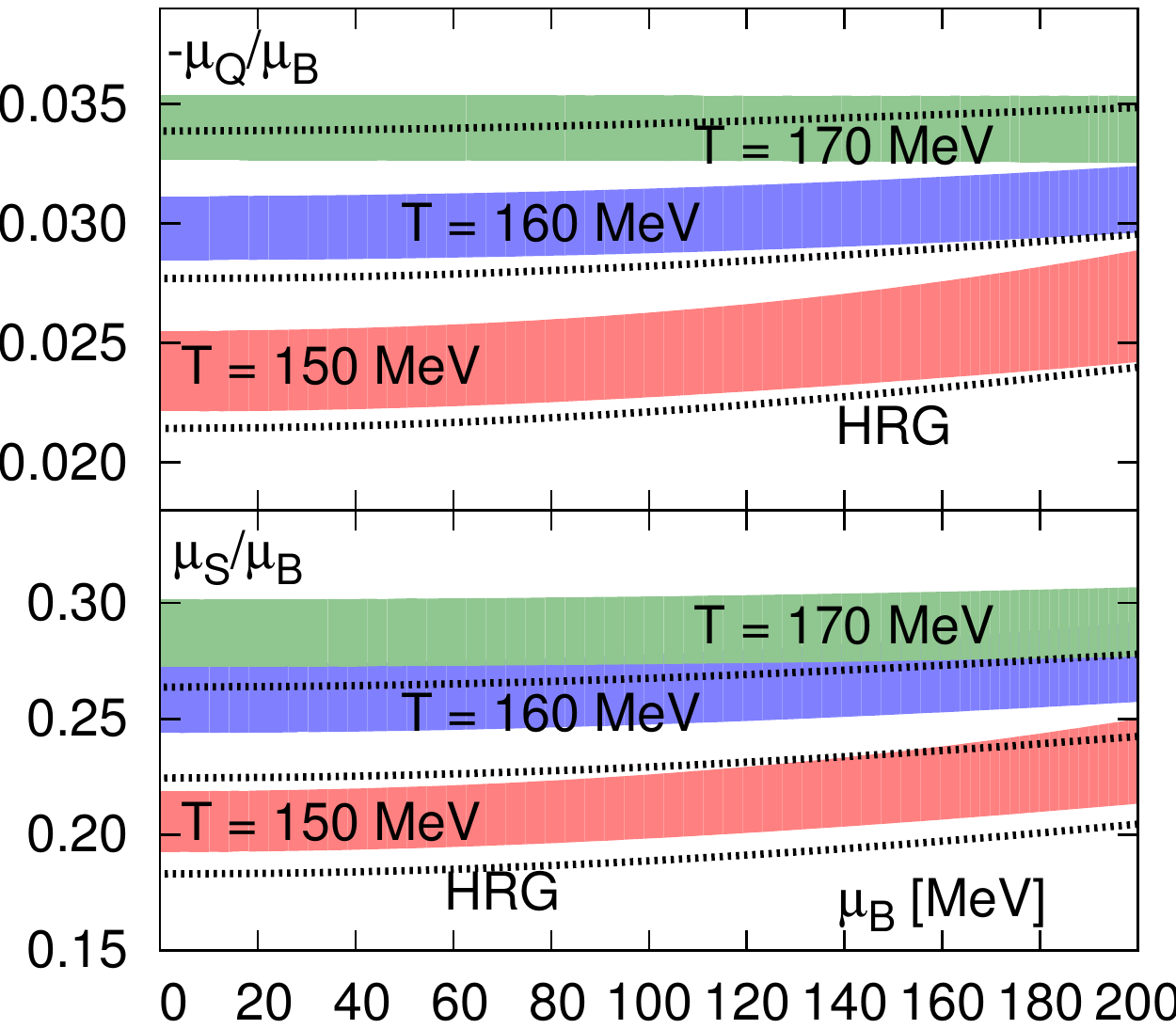} }
\caption{(a) LQCD results \cite{freeze} for the LO (top) and the  NLO (bottom) in
$\mu_B$ contributions for the electric charge chemical potential as a function of
temperature.  (b) Same as the previous panel, but for the strangeness chemical
potential. (c) Electric charge (top) and strangeness (bottom) chemical potential as a function of $\mu_B$ for the relevant temperature range $T=150-170$ MeV.}
\end{figure}

The freeze-out point is generally described by the freeze-out temperature $T^f$ and the freeze-out chemical potential $\vec{\mu}^f=(\mu_B^f,\mu_Q^f, \mu_S^f)$. In  heavy-ion collision these parameters are not independent. The conservation of strangeness and net electric charge during the fireball evolution constrains the corresponding chemical potentials to the values which govern the initial strangeness neutrality and net electric charge of the colliding nuclei. Assuming  spatial homogeneity and denoting the corresponding densities of the conserved charges ($X$) as $n_X$ these conditions are  expressed as \mbox{$\langle n_S\rangle=0$} and $\langle n_Q \rangle = r \langle n_B \rangle$. The initial fraction of charge particles $r$ is given by the number of protons divided by the number of protons and neutrons, $r=N_p/(N_p+N_n)$. In the following we will use  $r \approx 0.4$ as this well approximates the RHIC Au-Au and the LHC Pb-Pb collisions. 

By expanding $\lang n_X \rang$ using a Taylor series in powers of $(\mu_B,\mu_Q,\mu_S)$ up to $\mathcal{O}(\mu_X^3)$ and imposing the above constraints one can extract an expansion of $\mu_Q$ and $\mu_S$ in terms
of the $T$ and $\mu_B$ \cite{freeze}:
\begin{equation}
\mu_Q(T,\mu_B) = q_1(T)\mu_B + q_3(T)\mu_B^3 + \mathcal{O}(\mu_B^5)
\;, \quad 
\mu_S(T,\mu_B) = s_1(T)\mu_B + s_3(T)\mu_B^3 + \mathcal{O}(\mu_B^5)
\;. \label{eq:muQ-muS}
\end{equation}

In Fig.~\ref{fig:qi} and Fig.~\ref{fig:si} we show the results for the expansion coefficients $q_1(T)$ and $s_1(T)$ in Leading Order (LO) and the corresponding Next-to-Leading Order (NLO) corrections, $q_3(T)$ and $s_3(T)$. The NLO corrections are below 10\% in the relevant temperature range $160 \pm 10 \MeV$. For that range the  complete LO+NLO results for $\mu_Q$ and $\mu_S$ are shown in Fig.~\ref{fig:muQS}.
Note that for $T\approx157$ MeV the value for 
$\mu_S/\mu_B\approx0.24$ extracted from the LQCD data is in good agreement with the result  from statistical model
based fits of the strange baryons to anti-baryons ratios measured by the STAR
experiment \cite{zhao}. This backs up the assumption that strangeness neutrality is realized in the HIC and gives a first hint for the freeze-out temperature.

\section{\emph{Thermometer} and \emph{Baryometer} for heavy-ion collisions}

\begin{figure}
\subfigure[]{ \label{fig:R31Q} 
\includegraphics[height=0.24\textheight,width=0.48\textwidth]{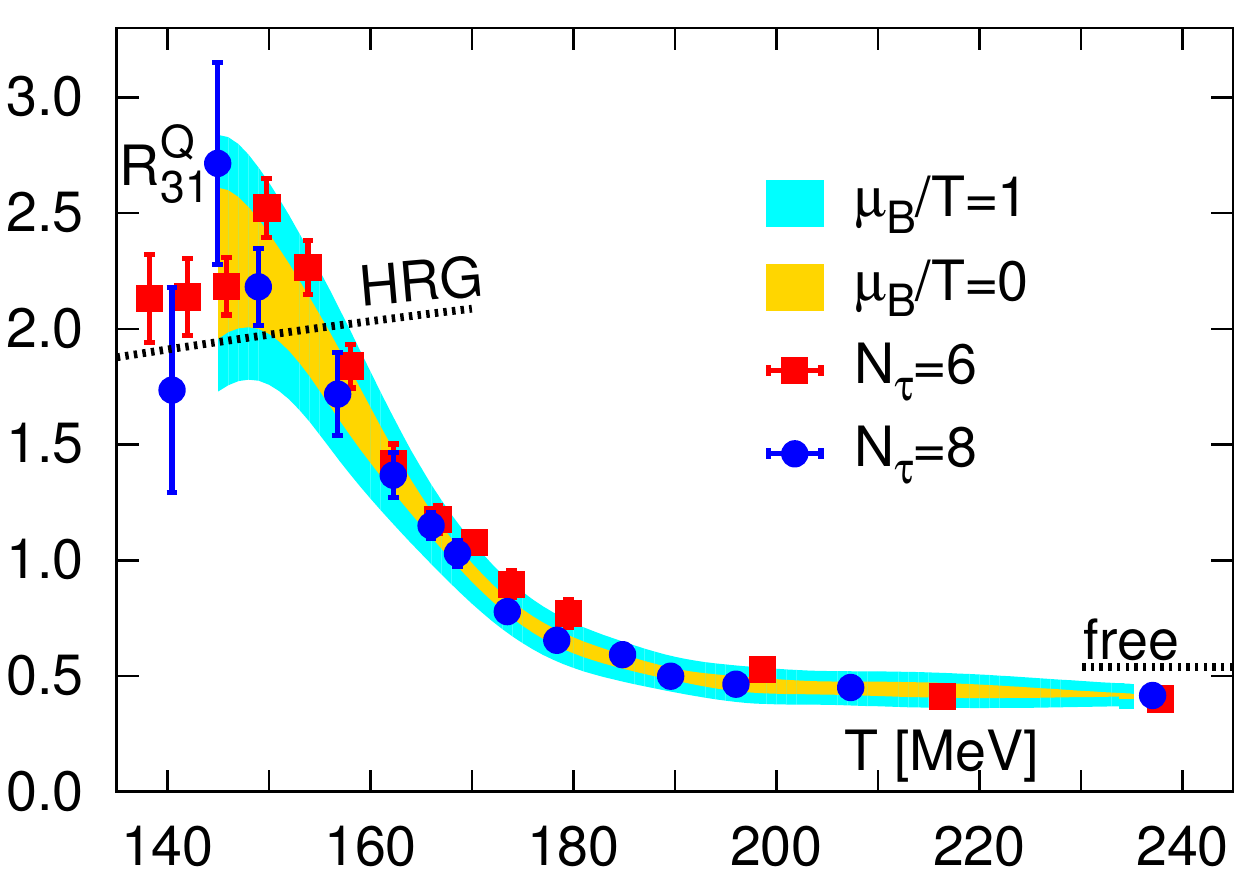} }
\subfigure[]{ \label{fig:R12Q} 
\includegraphics[height=0.24\textheight,width=0.48\textwidth]{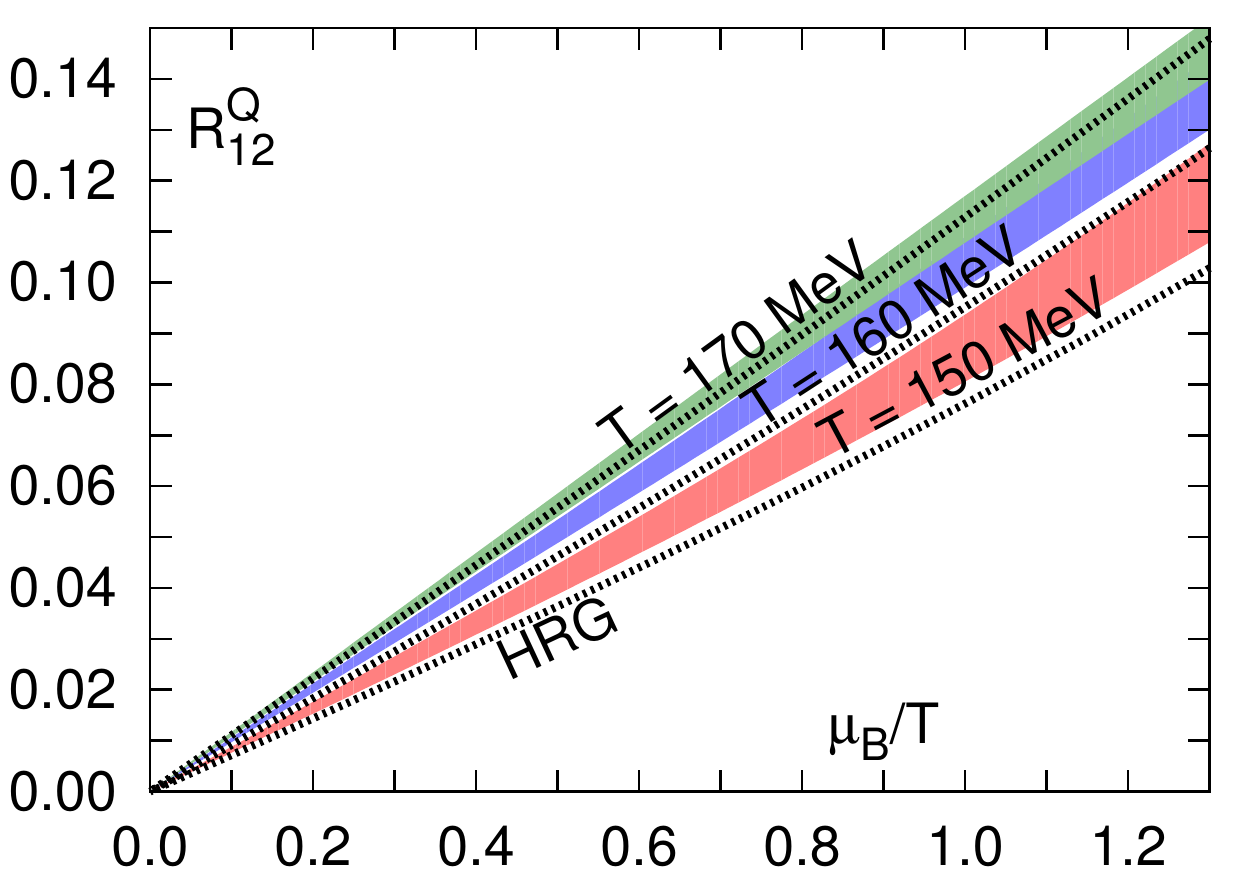} }
\caption{\label{fig:R31QR12Q}LQCD results \cite{freeze} for the \emph{thermometer} $R_{31}^Q$ (a) and the
\emph{baryometer} $R_{12}^Q$ (b) up to order $\mu_B^2$.}
\end{figure}

To obtain the freeze-out parameters we now need to determine the two remaining parameters, $T^f$ and $\mu_B^f$. Net electric charge fluctuations can be calculated from LQCD simulations and have -- in contrast to baryon number fluctuations -- also been measured in experiments at RHIC. To eliminate unknown explicit volume factors we will consider ratios of these fluctuations. At least two independent ratios are required to fix the freeze-out parameters. The use of more ratios, including  higher-order and/or baryon number or strangeness fluctuations, can provide an additional check of the thermodynamic consistency. Here we have chosen to work with

\beqa
R_{31}^Q &\equiv& \frac{\chi_3^Q(T,\mu_B)}{\chi_1^Q(T,\mu_B)} =
\frac{S_Q\sigma_Q^3}{M_Q} = R_{31}^{Q,0} + R_{31}^{Q,2} \mu_B^2 + \mathcal{O}(\mu_B^4)
\label{eq:R31Q} \\
R_{12}^Q &\equiv& \frac{\chi_1^Q(T,\mu_B)}{\chi_2^Q(T,\mu_B)} = \frac{M_Q}{\sigma_Q^2} 
= R_{12}^{Q,1} \mu_B + R_{12}^{Q,3} \mu_B^3 + \mathcal{O}(\mu_B^5)
\label{eq:R12Q} \;.
\eeqa

In LO $R_{31}^Q$ does not depend on the chemical potential and is therefore suitable as \emph{thermometer}. Once the temperature has been determined the in LO linear $\mu_B$-dependency of $R_{12}^Q$ motivates its choice as \emph{baryometer}. In Fig.~\ref{fig:R31QR12Q} we show our LQCD results for the ratios $R_{31}^Q$ and $R_{12}^Q$. For the latter we show three curves in the temperature range relevant for the heavy-ion collisions, $T=160\pm10 \MeV$. The NLO corrections are below 10\% and well controlled for $\mu_B \lesssim 200 \MeV$. This allows the use of the baryometer and thermometer for collision energies down to $\sqrt{S_{NN}} \gtrsim 19.6 \GeV$.

\begin{figure}[t!]
\subfigure[]{ \label{fig:R31Q-star-avg} 
\includegraphics[height=0.24\textheight,width=0.48\textwidth]{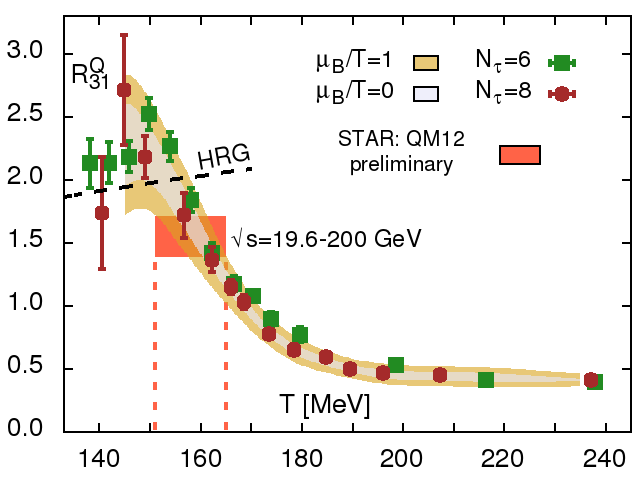} }
\subfigure[]{\label{fig:R12Q-phenix} 
\includegraphics[height=0.24\textheight,width=0.48\textwidth]{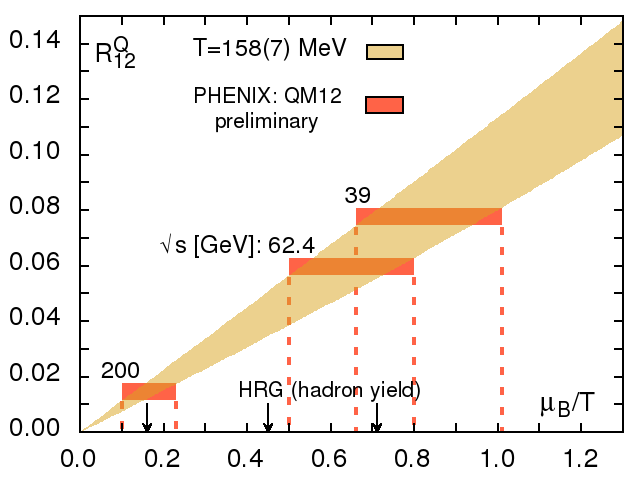} }
\subfigure[]{\label{fig:R12Q-star} 
\includegraphics[height=0.24\textheight,width=0.48\textwidth]{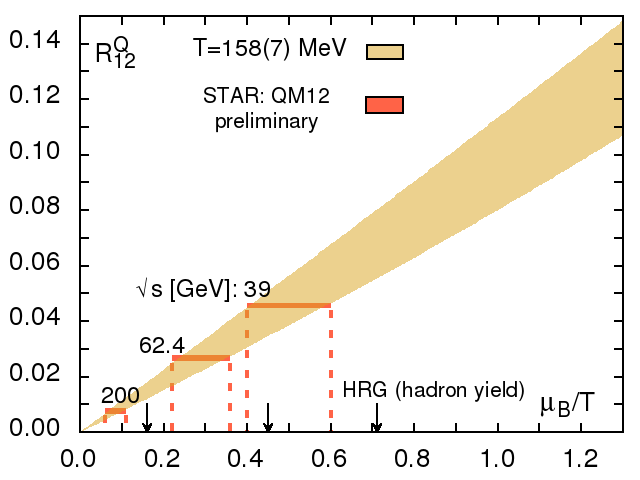} }
\subfigure[]{\label{fig:R12B-star} 
\includegraphics[height=0.24\textheight,width=0.48\textwidth]{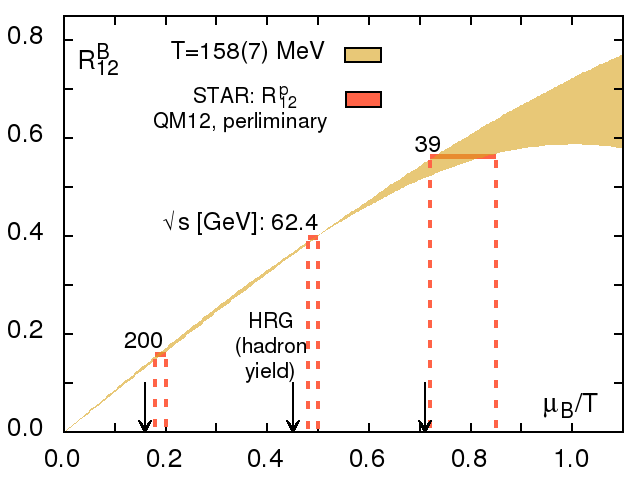} }

\caption{
(a) Comparisons between the LQCD results \cite{freeze} for the
\emph{thermometer} $R_{31}^Q$ and the preliminary STAR data \cite{star-charge} for
ratio $(S_Q \sigma_Q^3)/M_Q$ of the cumulants of the net electric charge fluctuation,
averaged over the energy range $\sqrt{s_{NN}}=19.6-200$ GeV. The overlap of the
experimental results with the LQCD calculations provides an estimate for the average
freeze-out temperature $T^f=158(7)$ MeV over $\sqrt{s}=19.6-200$ GeV.
(b) LQCD results \cite{freeze} for the \emph{baryometer} $R_{12}^Q$ as a function of $\mu_B/T$
compared with the preliminary PHENIX data \cite{phenix-charge} for $M_Q/\sigma_Q^2$
in the temperature range $T^f=158(7)$ MeV. The overlap regions of the experimentally
measured results with the LQCD calculations provide estimates for the freeze-out
chemical potential $\mu_B^f$ for a given $\sqrt{s_{NN}}$. The arrows indicate the
values of $\mu_B^f/T^f$ obtained from traditional statistical model fits to
experimentally measured hadron yields \cite{Cleymans:2005xv}.
(c) Same as (b) but with the preliminary STAR data
\cite{star-charge} for the ratio $M_Q/\sigma_Q^2$.
(d) Similar to (c) but here $R_{12}^B$ is compared with the preliminary STAR data \cite{star-proton} for the ratio $M_p/\sigma_p^2$ of the cumulants of net proton fluctuations.}
\end{figure}

To demonstrate the \emph{thermometer} and \emph{baryometer} we apply our method to preliminary data from STAR ~\cite{star-charge}. We compare our data for $R_{31}^Q$ with the experimental result for the corresponding ratio of cumulants of the net electric charge $(S_Q\sigma_Q^3)/M_Q$. This is shown in Fig.~\ref{fig:R31Q-star-avg}. The current experimental data still have large uncertainties and do not allow for an extraction of the $\sqrt{S_{NN}}$ dependency of the freeze-out temperature $T^f$~\cite{freezecpod}. To continue our analysis and extract the freeze-out chemical potential we therefore use the experimental data for $(S_Q\sigma_Q^3)/M_Q$ averaged over $\sqrt{S_{NN}} = 19.6 - 200 \GeV$. Note that for these collision energies also the traditional statistical model fits yield a very mild $\sqrt{S_{NN}}$-dependence of the freeze-out temperature. At even smaller collision energies the use of a $\mathcal{O}\left(\mu_B^3\right)$ Taylor series is no longer sound.
From our comparison we obtain an average freeze-out temperature $T^f = 158(7) \MeV$ for $\sqrt{S_{NN}} = 19.6 - 200 \GeV$. 

\begin{figure}[t!]
\begin{center}
\includegraphics[height=0.25\textheight,width=0.9\textwidth]{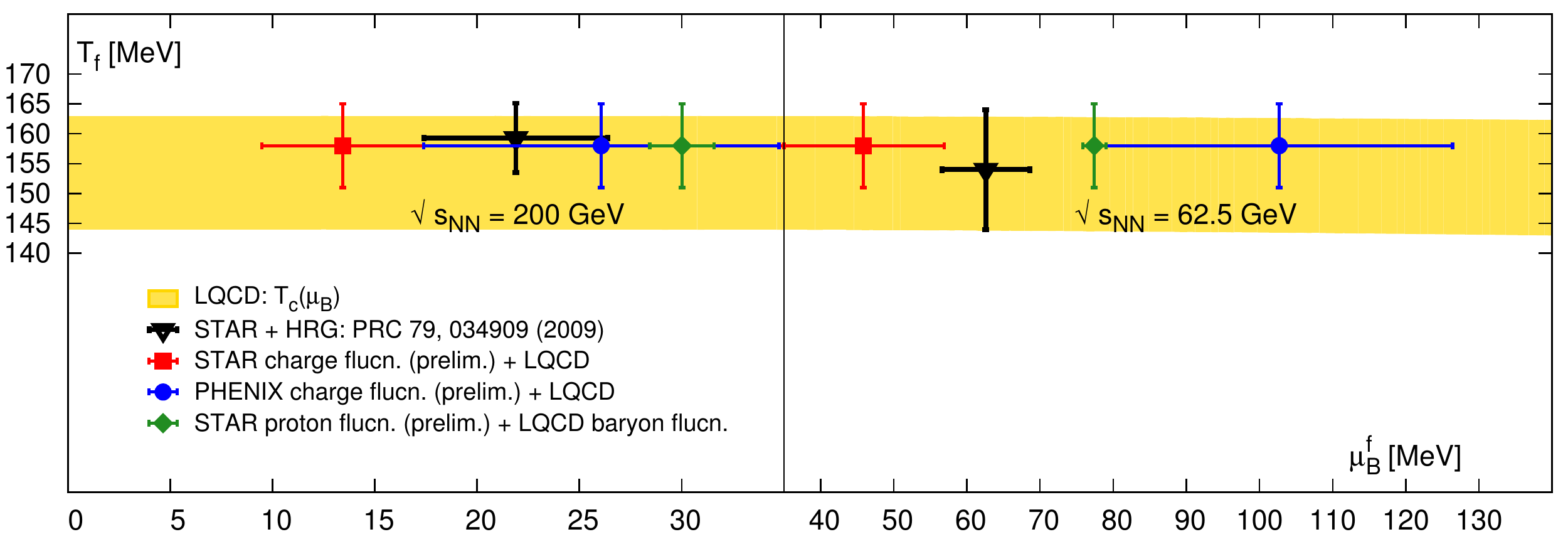}
\end{center}
\caption{Freeze-out temperatures $T^f$ and baryon chemical potentials $\mu_B^f$
obtained through direct comparisons between LQCD calculations and the preliminary
STAR and PHENIX data for cumulants of net charge and net proton fluctuations. The
shaded region indicate the LQCD results \cite{strange,Tc,curv} for the
chiral/deconfinement temperature $T_c$ as a function of the baryon chemical
potential.} 
\label{fig:tc-fo} 
\end{figure}

We can now determine also the freeze-out chemical potential from the comparison of the ratio $R_{12}^Q$ with the ratio of cumulants $M_Q/\sigma_Q^2$ measured at RHIC. In Fig.~\ref{fig:R12Q-phenix} we show the results for the preliminary data from the PHENIX experiment \cite{phenix-charge} for $\sqrt{S_{NN}}=200 \GeV, 62.4 \GeV, 39 \GeV$.  For each collision energy the freeze-out chemical potential is obtained as the overlap of the experimental result with the LQCD data on the $\mu_B/T$-axis. 
The black arrows indicate the freeze-out chemical potential obtained from statistical model fits~\cite{Cleymans:2005xv}. For the preliminary data available from the STAR experiment the comparison is shown in Fig.~\ref{fig:R12Q-star}.
The thermodynamic consistency of our approach can be checked either by using ratios using higher order fluctuations of conserved charges or by considering also baryon number and strangeness fluctuations, e.g., by using $R_{12}^B$ as \emph{baryometer}. Experimentally only proton number fluctuations have been measured which may be quantitatively quite different from the baryon number fluctuations~\cite{Bzdak:2012ab,Kitazawa}. Despite that we show for illustration the results using the alternative \emph{baryometer} $R_{12}^B$
and compare it with preliminary experimental data on proton fluctuations from STAR~\cite{star-proton}.
As can be seen from the figures the preliminary experimental data on net electric charge fluctuations from STAR and PHENIX yield quite different results for the freeze-out chemical potential. The proton data as third extraction method result in a third value for $\mu_B^f$. There is also no satisfactory agreement with the values from the traditional statistical model fits to hadron yields~\cite{star-fo}.
The situation is summarized in Fig.~\ref{fig:tc-fo} which shows the results for the freeze-out points for $\sqrt{S_{NN}}=200\GeV, 62.4 \GeV$ extracted using all four methods.
Until experimental uncertainties, differences in the extraction of the data between the experiments and the relation between baryon and proton number fluctuations have been ruled out this situation unlikely to improve. There is however also some positive news shown in Fig.~\ref{fig:tc-fo}: All freeze-out points from the different extractions are close to the chiral phase boundary for small values of the chemical potential $T_c(\mu_B)=\lrnd154(9)-[0.0066(7)/154(9)]\mu_B^2\rrnd$ MeV obtained from LQCD  \cite{strange,Tc,curv}. Hence the signals from the freeze-out might actually contain information about the critical behavior.

The LQCD based approach seems to be quite reliable. The Wuppertal-Budapest collaboration applied the discussed method using their lattice \cite{Borsanyi:2013hza} and the STAR data on electric charge fluctuations. Similar to our analysis they averaged $(S_Q\sigma_Q^3)/M_Q$ over several collision energies $\sqrt{S_{NN}}=27,39,62.4 \GeV$ and obtained $T^f \lesssim 157 \MeV$, close to our result $T^f = 158(7) \MeV$. Note however that we averaged over a larger range of collision energies ($\sqrt{S_{NN}}=19-200 \GeV$). Using the preliminary STAR data for net electric charge fluctuations they obtained $\mu_B^f=44(6) \MeV$ at  $\sqrt{S_{NN}}=62.4$, in agreement with our values.

Although the direct comparison of LQCD data with HIC experiments is fascinating and option some caution is required.  It is not certain whether the situation in HIC may be described within the grand canonical approach applied in LQCD simulation of QCD thermodynamics and the techniques used in the experimental analysis do not invalidate the approach~\cite{Bzdak:2012ab,Bzdak:2012an}. 
These issues are currently addressed in the experimental analysis \cite{star-charge,phenix-charge,star-proton}. Until these are resolved and better experimental are available the experimental data remain the limiting factor in the approach.

\section*{Acknowledgement}
MW acknowledges support by the BMBF under grant 05P12PBCTA, the numerical calculations have been performed using the USQCD
GPU-clusters at JLab, the Bielefeld GPU cluster and the NYBlue at the NYCCS.


\end{document}